# Modeling the amorphous structure of a mechanically alloyed $Ti_{50}Ni_{25}Cu_{25}$ alloy using anomalous wide-angle x-ray scattering and reverse Monte Carlo simulations

J.C. de Lima[a,*], C.M. Poffo[b], S.M. Souza[c], K.D. Machado[d], D.M. Trichês[c], T.A. Grandi[a], R.S. de Biasi[e]

[a]*Departamento de Física, Universidade Federal de Santa Catarina, Campus Universitário Trindade, S/N, C.P. 476, 88040-900 Florianópolis, Santa Catarina, Brazil*

[b]*Departamento de Engenharia Mecânica, Universidade Federal de Santa Catarina, Campus Universitário Trindade, S/N, C.P. 476, 88040-900 Florianópolis, Santa Catarina, Brazil*

[c]*Departamento de Física, Universidade Federal do Amazonas, 3000 Japiim, 69077-000 Manaus, Amazonas, Brazil.*

[d]*Departamento de Física, Centro Politécnico, Universidade Federal do Paraná, 81531-990, Curitiba, Paraná, Brazil*

[e]*Seção de Engenharia Mecânica e de Materiais, Instituto Militar de Engenharia, 22290-270 Rio de Janeiro, RJ, Brazil*

**A B S T R A C T**

An amorphous $Ti_{50}Ni_{25}Cu_{25}$ alloy was produced by 19 h of mechanical alloying. Anomalous wide angle x-ray scattering data were collected at six energies and six total scattering factors were obtained. By considering the data collected at two energies close to the Ni and Cu K edges, two differential anomalous scattering factors around the Ni and Cu atoms were obtained, showing the chemical environments around these atoms are different. The eight factors were used as input data to the reverse Monte Carlo method used to compute the partial structure factors $S_{Ti-Ti}(K)$, $S_{Ti-Cu}(K)$, $S_{Ti-Ni}(K)$, $S_{Cu-Cu}(K)$, $S_{Cu-Ni}(K)$ and $S_{Ni-Ni}(K)$. From their Fourier transformation, the partial pair distribution functions $G_{Ti-Ti}(r)$, $G_{Ti-Cu}(r)$, $G_{Ti-Ni}(r)$, $G_{Cu-Cu}(r)$, $G_{Cu-Ni}(r)$ and $G_{Ni-Ni}(r)$ were obtained, and the coordination numbers and interatomic atomic distances for the first neighbors were determined.

*Corresponding author. Tel.:+55 48 37212847; fax:+55 48 37219946.
E-mail address:fsc1jcd@fisica.ufsc.br (J.C. de Lima).

# 1. Introduction

Alloys with shape memory effects (SME) based on TiNi alloys have been used in different areas of science and technology, such as electronics, medicine and the space industry [1]. Among them, the $Ti_{50}Ni_{25}Cu_{25}$ has been widely studied, but knowledge about its structure and properties is still incomplete. In part, this is due to the fact that this alloy is produced in the amorphous state and the crystalline $Ti_2NiCu$ phase is obtained via crystallization [2]. Knowledge of its structure in the amorphous state seems to be mandatory for a better understanding of the relationship between properties and microstructure.

The structure of an amorphous alloy containing $n$ constituents is described by $n(n+1)/2$ partial pair correlation functions $G_{ij}(r)$, which are related to the partial structure factors $S_{ij}(K)$ through a Fourier transformation. Here, $|K| = 4\pi(\sin\theta)/\lambda$ is the transferred wave vector. The total structure factor $S(K)$, which can be derived from scattering measurements, is a weighted sum of these $n(n+1)/2$ $S_{ij}(K)$ factors [3,4]. Thus, in order to determine the $n(n+1)/2$ $S_{ij}(K)$ factors, at least the same number of independent $S(K)$ factors are needed. Usually, the isomorphous substitution and isotope substitution methods [5,6] have been used to obtain these factors.

With the development of synchrotron radiation sources, anomalous wide angle x-ray scattering (AWAXS) and differential anomalous scattering (DAS) techniques became available for structural study of multicomponent disordered materials. AWAXS utilizes an incident radiation that is tuned close to an atomic absorption edge so that it interacts resonantly with the electrons of that particular atom. The atomic scattering factor, $f(K,E) = f_0(K) + f´(E) + if´´(E)$, of each chemical component can therefore be varied individually and the chemical environment about each component in the material can be investigated. Thus, in the case of an amorphous alloy containing $n(n+1)/2$ constituents, the $n(n+1)/2$ independent $S(K)$ factors can be obtained from a single sample. However, the matrix formed by their weights is ill-conditioned, compromising the determination of the $n(n+1)/2$ $S_{ij}(K)$ factors.

Fuoss and co-workers [7,8] tried to overcome this difficulty by implementing the differential anomalous scattering (DAS) approach, which was proposed by Schevchik [9,10]. The DAS approach consists of taking the difference between the scattering patterns measured at two incident photon energies just below the edge of a particular atom, so that all correlations not involving this atom subtract out, since only the atomic scattering factor of this atom changes appreciably. Later, de Lima et al. [11], following a suggestion made by Munro [12], combined the differential scattering factors $DSF(K)$ and the $S(K)$ factors to obtain the three $S_{Ni-Ni}(K)$, $S_{Ni-Zr}(K)$ and $S_{Zr-Zr}(K)$ factors for the amorphous $Ni_2Zr$ alloy. They observed that this combination reduces the conditioning number of the matrix formed by the weights of these factors, allowing more stable values of $S_{ij}(K)$ to be obtained.

The reverse Monte Carlo (RMC) simulation technique [13–16] has been successfully used for structural modeling of amorphous structures. One or more $S(K)$ factors or their Fourier transformations, named as total pair correlation functions $G(r)$, can be used as input data. Recently, de Lima et al. [3,4] reported the determination of the $S_{Ni-Ni}(K)$, $S_{Ni-Zr}(K)$ and $S_{Zr-Zr}(K)$ factors for amorphous $NiZr_2$ and $NiZr_3$ alloys by making a combination of AWAXS, DAS and RMC simulation techniques. These excellent results have motivated us to apply the RMC simulations to a ternary amorphous $Ti_{50}Ni_{25}Cu_{25}$ alloy produced by mechanical alloying (MA). Thus, the aim of this paper is to report the partial $S_{ij}(K)$ factors obtained considering the $S(K)$ and Ni- and Cu-$DSF(K)$ factors as input data.

## 2. Experimental procedure

### 2.1 Sample preparation

A stoichiometric ternary $Ti_{50}Ni_{25}Cu_{25}$ mixture of elemental powders of Ti (Alfa Aesar, 60-100 mesh, purity 99.9%), Ni (Alfa Aesar, purity 99.9%, 2.2-3 $\mu$m) and Cu (Alfa Aesar, 19 $\mu$m, purity 99.9%) was sealed together with several steel balls 11.0 mm in diameter into a cylindrical steel vial under argon atmosphere. The ball-to-powder weight ratio was 4:1. The vial was mounted on a SPEX mixer/mill, model 8000. The temperature was kept close to room temperature by a ventilation system. After 19 h of milling, the measured XRD pattern showed broad halos characteristic of amorphous materials and the milling process was interrupted. No peaks of elemental Ti, Ni, Cu or any crystalline phase were observed. The XRD patterns were acquired using a Miniflex Rigaku powder diffractometer, equipped with Cu$K_\alpha$ radiation ($\lambda = 1.5418$ Å).

### 2.2 AWAXS measurements and data analysis

AWAXS scattering experiments are performed with synchrotron radiation sources using a two-circle diffractometer in the vertical plane equipped with a channel-cut Si(220) single crystal monochromator and a Si:Li energy-sensitive detector. The pulses are processed by a multichannel analyzer. This energy-sensitive detector is able to discriminate the large $K_\alpha$ resonant Raman or fluorescence signal when the incident photon energy is tuned close to the K edges, but cannot distinguish the $K_\beta$ fluorescence signal and inelastic scattering (Compton) from the elastically scattered intensity. To minimize the air scattering at low $K$-range, the sample is sealed, under vacuum, into a cell containing a large kapton window fixed around the diffractometer horizontal rotation. The content of $K_\beta$ fluorescence signal is estimated using theoretical $K_\beta/K_\alpha$ ratio.

Details on the AWAXS apparatus and data analysis are described in Refs. [3,4,11,17 and 18].

In this study, the AWAXS measurements were performed at the DB12A (XRD1) beamline of the Brazilian Synchrotron Light Laboratory (LNLS), which is equipped with a sagittal focusing double crystal Si(111) monochromator, a Huber diffractometer with a new arm to improve the stability of the analyzing crystals, slits, detector and a cyberstar scintillation detector [19,20]. The energy and average current of the storage ring were 1.37 GeV and 150 mA, respectively. No vacuum cell around the sample was used to minimize the air scattering. Due to the use of a scintillation detector, a graphite analyzer ($d$ = 3.3585 Å) was used in the secondary beam to suppress the Ni and Cu $K_\alpha$ fluorescence signal. However, its mosaic spread (0.4-0.5°) did not permit to distinguish the $K_\beta$ fluorescence signal and inelastic scattered intensity from the elastically scattered intensity. As the theoretical $K_\beta/K_\alpha$ ratio is commonly used to estimate the content of $K_\beta$ fluorescence signal overlapped to the scattered intensity, the Ni and Cu $K_\alpha$ fluorescence signals were measured by tuning the graphite analyzer at the angles θ = 14.301° and 13.269°, respectively. Unfortunately, broad halos of low intensity on the Ni and Cu $K_\alpha$ fluorescence signals were observed, making impossible their use. These halos were located at the same $K$ values where on the measured scattered intensities halos are observed. Thus, an approach to subtract Ni and Cu $K_\beta$ fluorescence and air scattering contributions to measured scattered intensity was developed and it will be described in detail in another section.

The elastically plus inelastically scattered intensities remaining after subtraction of air scattering and $K_\beta$ fluorescence signal were corrected for reabsorption effects being then put on a per-atom scale and the inelastic scattered intensity subtracted [21]. Due to the diffractometer characteristics (acquisition of data in the vertical plane), the polarization correction was disregarded. The inelastic scattered intensity was calculated according to the analytic approximation given by Pálinkas [22].

## 3. Determination of the real and imaginary parts of the atomic scattering factor

In order to interpret the scattering data correctly, the real and the imaginary parts $f´$ and $f´´$ of the atomic scattering factor were determined following a procedure described by Dreier et al. [23] and used by us in other papers [3,4,11]. For this, x-ray absorption (XAS) coefficients were measured at LNLS near Ni and Cu K edges on the sample and $f´´$ was calculated using the optical theorem. Outside the region of measurement, theoretical values of $f´´$ taken from a table compiled by Sasaki [24] were used to extend the experimental data set over a larger energy range and $f´$ was calculated using the Kramers-Kronig relation, as illustrated in Figs. 1-3. For the measurements away from the K edges, the $f´$ and $f´´$ values given in table compiled by Sasaki were used. The resulting values for incident photon energies 8333 eV and 8979 eV are listed in Table I together with Sasaki values. The atomic scattering factor away from the K edge $f_0(K)$ of neutral Ti, Ni and Cu atoms were calculating according to the analytic function given by Cromer and Mann [25].

## 4. Approach to subtract the air scattering and the content of $K_\beta$ fluorescence signal of the measured scattered intensity

With the exception of an incident photon energy of 8233 eV, where no Ni and Cu $K_\beta$ and $K_\alpha$ fluorescence signals were generated, for all other incident photon energies listed in Table 1 the measured scattered intensity is the sum of elastic, inelastic intensities, air scattering and Ni or Cu $K_\beta$ fluorescence signal. Thus, to explore the experimental data the elastically scattered intensity, it must be isolated. For this, an approach to subtract the air scattering and the $K_\beta$ fluorescence signal from the measured scattered intensity was developed. The air scattering contribution was measured by removing the sample and sample holder. It has the shape of an exponential decay and is significant up to $K \approx 2$ Å$^{-1}$. Two procedures were tried to subtract its contribution: that described in Ref. [26], and another in which the air scattering was fitted to an exponential decay. For the fitting, the baseline tools present in the Origin software [27] was used. The results were similar, and since the last is easier, it was kept.

Experimentally, $K_\alpha$ and $K_\beta$ fluorescence signals increase with increasing $K$ values and have no peaks. Considering the impossibility of using the measured Ni and Cu $K_\alpha$ fluorescence signals due to the broad halos of low intensity overlapped to them, it was assumed that the content of $K_\beta$ fluorescence signal overlapped to the measured scattered intensity can be represented by an arbitrary function $F(K)$, which can be drawn by using the baseline tools present in the Origin software [27]. Fig. 4 shows the fits of air scattering (exponential decay) and $K_\beta$ fluorescence signal that were subtracted from the measured scattered intensity at the energy 8333 eV, leaving only the sum of elastic and inelastic intensities offset count to zero. The missing intensity (offset count from zero) was found by considering the measured scattered intensity at the energy 8233 eV, which is offset count from zero by determined value. For this energy, the selected number of counts per point was the smallest. At the DB12A beamline, the software used to control the monochromator, the diffractometer and to record the measured scattered intensity uses the number of counts per point to take into account the time decrease of the beam. Depending of selected energy and average current of the storage ring, different numbers of counts per point were selected, resulting that scattered intensities with different content of air scattering and $K_\beta$ fluorescence signal were recorded. After subtracting air scattering and $K_\beta$ fluorescence signal, the missing intensity (the offset count from zero) was obtained by assuming that it is proportional to that present in the measured scattered intensity at the energy 8233 eV. The constant of proportionality was calculated considering the ratio between the integrated measured scattered intensity at selected energy and the integrated one measured at the energy of 8233 eV. This approach has successfully applied for all measured scattered intensities at the energies higher than 8233 eV, as will be shown in the next sections.

## 5. Results and discussion

### 5.1 Total structure factors, differential structure factors, and differential distribution functions

The sum of elastically and inelastically scattered intensities was obtained after subtraction of air scattering and $K_\beta$ fluorescence signal. It was corrected for reabsorption effects and put on a per-atom scale following the procedure described in Ref. [21] and the inelastic scattered intensity subtracted. Fig. 5 shows the elastically scattered intensities put on a per-atom scale together with the mean-square scattering factors $<f^2(K, E)> = x_{Ti} f^2_{Ti}(K,E) + x_{Ni} f^2_{Ni}(K,E) + x_{Cu} f^2_{Cu}(K,E)$, where $x_i$ is the concentration, for the energies listed in Table 1. The procedure to derive the total structure factor $S(K)$ from the elastically scattered intensities put on a per-atom scale as well as to obtain the reduced total distribution function $\gamma(r)$ [$\gamma(r) = 4\pi\rho_0 r[G(r)-1]$, the total pair distribution function $G(r)$ and the total radial distribution function $RDF(r)$ [$RDF(r) = 4\pi\rho_0 r^2 G(r)$] is detailed in Refs. [3,4,21] and will not be repeated here. $\rho_0$ is the atomic number density (atoms/Å$^3$). The $\gamma(r)$ function is related to the $S(K)$ function through a Fourier transform.

Fig. 6 shows the $S(K)$ function for the incident photon energies listed in Table 1. Comparison shows that up to $K = 4$ Å$^{-1}$ they are similar, whereas for higher $K$ values they show significant differences. Since $S(K)$ is a weighted sum of $S_{ij}(K)$, we calculated the weights $W_{ij}(K)$ for the $S_{Ti-Ti}(K)$, $S_{Ti-Cu}(K)$, $S_{Ti-Ni}(K)$, $S_{Cu-Cu}(K)$, $S_{Cu-Ni}(K)$ and $S_{Ni-Ni}(K)$ factors. For $K = 2.93$ Å$^{-1}$, the contributions to the $S(K)$ function at 8330 eV are about 23%, 30%, 20%, 10%, 13% and 4%, respectively, while to $S(K)$ factor at 8979 eV are about 23%, 21%, 29%, 5%, 13% and 9%, respectively. Due to the small contributions to $S(K)$, the determination of $S_{Cu-Cu}(K)$, $S_{Cu-Ni}(K)$ and $S_{Ni-Ni}(K)$ is more difficult than the determination of $S_{Ti-Ti}(K)$, $S_{Ti-Cu}(K)$ and $S_{Ti-Ni}(K)$. At 9077 eV, $S(K)$ achieved the highest value $K_{max}$ (8 Å$^{-1}$), introducing a fictitious breadth $\Delta r \approx 0.475$ Å [$\Delta r = 3.8/K_{max}$] to the peaks of the $\gamma(r)$, $G(r)$ and $RDF(r)$ functions. Fig. 6 shows that all the $S(K)$ factors have a main halo at about $K \approx 2.93$ Å$^{-1}$. The average interatomic distance corresponding to this halo may be estimated through the Ehrenfest relation $r = \lambda/E\sin\theta = 4\pi/EK$. If the structure dependent constant $E$ is taken as 1.671 [28], an $r$ value of 2.57 Å is obtained.

Fig. 7 shows the $\gamma(r)$ function corresponding to the $S(K)$ factors shown in Fig. 6. For $r > 10$ Å, the oscillations are very weak, and therefore, they are shown only up to $r = 10$ Å. The straight line $\gamma(r) = 4\pi\rho_0 r$ is shown together with the $\gamma(r)$ function for an energy of 8233 eV. The oscillations are related to the small $K_{max}$ value achieved in $S(K)$ factors and the $\gamma(r)$ functions must oscillate about the straight line, as shown in Fig. 7. The slope of the $\gamma(r)$ function is equal to $4\pi\rho_0$ and the density of alloy can be calculated. For the amorphous Ti$_{50}$Ni$_{25}$Cu$_{25}$ alloy, a value of $\rho_0 = 0.06818$ atoms/Å$^3$ (6.1714 g/cm$^3$) was obtained. The JCPDS Database [29] gives a density between 6.481 and 6.743 gr/cm$^3$ for TiNi, while for the TiCu the density is between 6.509 and 6.574 g/cm$^3$. From this figure, one can see that the first neighbor shell, with an average distance of $r = 2.70$ Å, is well isolated, while the second one, between $r = 3.50$ and 6 Å, is broad and splits into two sub-shells with increasing photon energy.

The formalism of the differential structure factor $DSF(K,E_m,E_n)$ about a specific atom is described in Refs. [3,4] and will not be repeated here. Fig. 8 shows the Ni-

$DSF(K)$ and Cu-$DSF(K)$ factors obtained from the difference between the elastic scattered intensities in a per-atom scale measured at 8333 and 8433 eV and at 8979 and 9077 eV, respectively. Due to the high noise level present in these functions, they were smoothed using the smoothing tool (FFT filter considering 7 points) available in the Origin software [27]. All physical information present in these functions was preserved after smoothing, as shown in this figure. The Ni-$DSF(K)$ and Cu-$DSF(K)$ functions describe the chemical environments about the Ni and Cu atoms in the amorphous $Ti_{50}Ni_{25}Cu_{25}$ alloy, and one can see that they are different, as shown in Fig. 8. For example, the main halo at about $K = 3$ Å$^{-1}$ is better defined in the Ni-$DSF(K)$ factor, meaning that the chemical bonds involving the Ni atoms are more ordered than those involving the Cu atoms. Since Ni-$DSF(K)$ factor is a weighted sum of the $S_{ij}(K)$ factors, the weights $Wij(K)$ for the $S_{Ti\text{-}Ti}(K)$, $S_{Ti\text{-}Cu}(K)$, $S_{Ti\text{-}Ni}(K)$, $S_{Cu\text{-}Cu}(K)$, $S_{Cu\text{-}Ni}(K)$ and $S_{Ni\text{-}Ni}(K)$ factors were calculated. At $K = 2.93$ Å$^{-1}$, their contributions to Ni-$DSF(K)$ are about 0.2%, 1%, 48%, 1%, 30% and 24%, respectively. Thus, the Ni-$DSF(K)$ factor is a weighted sum of $S_{Ti\text{-}Ni}(K)$, $S_{Cu\text{-}Ni}(K)$ and $S_{Ni\text{-}Ni}(K)$ factors. For the Cu-$DSF(K)$ factor, they are about 0.05%, 45%, 1%, 24%, 29% and 0.8%, respectively. Thus, the Cu-$DSF(K)$ factor is a weighted sum of $S_{Ti\text{-}Cu}(K)$, $S_{Cu\text{-}Cu}(K)$ and $S_{Cu\text{-}Ni}(K)$ factors. The weights of the $S_{Ni\text{-}Ni}(K)$ and $S_{Cu\text{-}Cu}(K)$ factors suggest that more stable factors can be obtained considering the six $S(K)$ plus the Ni-$DSF(K)$ and Cu-$DSF(K)$ factors as input data for the RMC simulations. The differential distribution function $DDF(r)$ is related to $DSF(K,E_m,E_n)$ through a Fourier transformation. Fig. 9 shows the $DDF(r)$ functions corresponding to the Ni-$DSF(K)$ and Cu-$DSF(K)$ factors shown in Fig. 8, and they corroborate the fact that the chemical environments about the Ni and Cu atoms are different in this amorphous alloy.

*5.2 Partial structure factors obtained from the reverse Monte Carlo simulations*

The basic idea and the algorithm of the standard reverse Monte Carlo (RMC) method are described elsewhere [13–16] and its application to different materials is documented in the literature. A brief summary is described in Refs. [3,4] and will not be repeated here.

For the RMC simulations, the density value $\rho_0 = 0.06818$ atoms/Å$^3$ (6.1714 g/cm$^3$) and 5000 atoms (2500 Ti, 1250 Ni and 1250 Cu) were used to generate an initial random configuration, without unreasonably short interatomic distances, in a cubic box of edge $L = 42$ Å. The density $\rho_0$ of the amorphous $Ti_{50}Ni_{25}Cu_{25}$ alloy is an important input parameter of the RMC simulations. Since it is not easy to determine $\rho_0$ experimentally, because the alloy is an amorphous powder, the procedure described in Ref. [30] was used to obtain the best density value.

It is well known that the $G_{ij}(r)$ functions have the first neighbor shells well represented by one or more Gaussian function. With the exception of pre-peaks, which are associated with intermediate range order, those located before the first shell have no physical meaning. These features should be pursued in any method used for modeling the atomic structure of amorphous materials. We assumed cutoff distances $\Delta_{ij}$ in the RMC simulations to act as constraints on the short-range structure, and these were carefully investigated. In the absence of a direct Fourier transform, there are no criteria for choosing them; however, some physical considerations about the types of atoms as well as their contents in the alloy may be useful. The atomic radii of Ti, Ni and Cu atoms are 1.45 Å, 1.25 Å and 1.28 Å, respectively; the Ni and Cu contents in the amorphous $Ti_{50}Ni_{25}Cu_{25}$ alloy are 25 at.% (diluted alloy). Thus, the value of $\Delta_{Ni-Ni}$ and $\Delta_{Cu-Cu}$ may be greater than those involving Ti atom ($\Delta_{Ti-Ti}$, $\Delta_{Ti-Ni}$ and ($\Delta_{Ti-Cu}$). Based on these considerations, several sets of cutoff distances $\Delta_{ij}$ were examined. Each $\Delta_{ij}$ set was introduced in the initial random configuration before submitting it to a process to maximize the amount of disorder (entropy). This process is well described in the RMC manual. In order to prevent the presence of spurious artifacts in $S_{ij}(K)$ and $G_{ij}(r)$ that could lead to misinterpretations, the $S(K)$ factors were smoothed using the smoothing tool (FFT filter and considering 5 points) available in the Origin software [27]. All physical information present in the original $S(K)$ factors was preserved. After this, the configuration, the six smoothed $S(K)$ factors listed in Table 1 and the smoothed Ni-$DSF(K)$ and Cu-$DSF(K)$ factors were used for the RMC simulations.

The best simulations were achieved considering the cutoff distances $\Delta_{Ti-Ti} = 2.15$ Å and $\Delta_{Ti-Ni} = \Delta_{Ti-Cu} = \Delta_{Ni-Ni} = \Delta_{Cu-Cu} = 2.0$ Å. A second input data set formed by the smoothed $S(K)$ factors for the energies 8333, 8433, 8877 and 8979 eV was also considered. Fig. 10 shows the experimental (open circle curves) and simulated (solid lines) $S(K)$ factors for the second input data set, and one can see a good agreement between them.

Figs. 11 and 12 show the $S_{Ti-Ti}(K)$, $S_{Ti-Cu}(K)$, $S_{Ti-Ni}(K)$, $S_{Cu-Cu}(K)$, $S_{Cu-Ni}(K)$ and $S_{Ni-Ni}(K)$ factors and $G_{Ti-Ti}(r)$, $G_{Ti-Cu}(r)$, $G_{Ti-Ni}(r)$, $G_{Cu-Cu}(r)$, $G_{Cu-Ni}(r)$ and $G_{Ni-Ni}(r)$ obtained from the RMC simulations considering the both input data sets. The open circle curves are the $S_{ij}(K)$ factors obtained using the six $S(K)$ plus the two $DSF(K)$ factors, while the dark gray solid curves are the $S_{ij}(K)$ factors obtained using the four $S(K)$ factors. In Fig. 11, one can see a good agreement among the $S_{ij}(K)$ factors obtained using both input data sets, despite the small contribution of the $S_{Cu-Cu}(K)$ and $S_{Ni-Ni}(K)$ factors to $S(K)$. It is interesting to note that the main halos of $S_{ij}(K)$ associated with the pair correlations containing Cu atoms are broader than the other ones, suggesting that these chemical bonds are more disordered than those containing Ni and Ti atoms, as was already pointed out. Other feature involving this atom is observed in the $S_{Cu-Cu}(K)$ factor that shows the main halo partially separated into two other ones. In Fig. 12, one can see a good agreement between $G_{ij}(r)$ functions obtained for both input data sets. It is interesting to note that the first neighbor shells of $G_{Ti-Cu}(r)$, $G_{Cu-Cu}(r)$ and $G_{Cu-Ni}(r)$ are broad and asymmetrical, corroborating the presence of a significant chemical disorder in these pair correlation functions. On the other hand, the first neighbor shells of the $G_{Ti-Ti}(r)$, $G_{Ti-Ni}(r)$, and $G_{Ni-Ni}(r)$ functions are narrow, symmetrical and well isolated, indicating the presence of chemical ordering in these chemical bonds.

The interatomic distances for the first neighbors are those corresponding to the first maxima of the $G_{ij}(r)$ functions, and the coordination numbers were calculated from the partial radial distribution function $RDF_{ij}(r)$ [$RDF_{ij}(r) = 4\pi\rho_0 C_j r^2 G_{ij}(r)$] considering the partial pair distribution functions $G_{ij}(r)$ shown in Fig. 12. The upper limits were taken as being the minima located between the first and second neighbor shells. The values are 2.87 ± 0.08 Ni-Ni pairs at 2.71 Å, 3.04 ± 0.05 Cu-Ni pairs at 2.64 ± 0.07 Å, 3.05 ± 0.21 Ti-Ni pairs at 2.71 Å, 6.32 ± 0.12 Ti-Ti pairs at 2.71 Å, 3.32 Cu-Cu pairs at 2.64 ± 0.07 Å and 3.31 ± 0.01 Ti-Cu pairs at 2.71 Å. Louzguine and Inoue [2] studied the crystallization behavior of the amorphous $Ti_{50}Ni_{25}Cu_{25}$ alloy. A single stage polymorphic-type transformation of the amorphous phase forming a $Ti_2CuNi$ crystalline phase was observed. Those researchers indexed the XRD pattern of the $Ti_2NiCu$ phase to a cubic lattice (S.G. Pm-3m) with a lattice parameter $a = 3.047$ Å. According to them, the lattice parameter of $Ti_2NiCu$ is very close to that of the cubic TiNi phase, indicating that this phase is a solid solution of Cu in TiNi phase with Cu atom replacing Ni atoms located at the corners of the unit cube. At ambient conditions, the TiNi phase crystallizes in a cubic structure (S.G. Pm-3m) with lattice parameter $a = 2.972$ Å. The Ni and Ti atoms occupy the 1a (0 0 0) and 1b (1/2 1/2 1/2) Wyckoff sites. These structural data were used together with the Crystal Office 98 software [31] to build the

3D structure, and from 6 Ni-Ni pairs at 3.010 Å, 6 Ti-Ti pairs at 3.010 Å and 8 Ti-Ni pairs at 2.607 Å were obtained. In another paper [30], we modeled the amorphous structure of the $Ni_{46}Ti_{54}$ alloy through the RMC simulations. From the final atomic configuration 5.5 Ni-Ni pairs at 2.67 Å, 6.5 Ti-Ti pairs at 2.71 Å and 5.7 Ti-Ni pairs at 2.63 Å were obtained. These studies show that the number of first neighbor Ni-Ni and Ti-Ti pairs are similar in both amorphous $Ni_{46}Ti_{54}$ and crystalline TiNi phases but the interatomic distances are smaller in amorphous phase. The number of first neighbor Ti-Ni pairs is greater in crystalline TiNi phase but the interatomic distances are similar.

In this study, the coordination numbers obtained for the first neighbors suggest that in the amorphous $Ti_{50}Ni_{25}Cu_{25}$ phase the Cu atom replaces Ni one as the crystalline cubic $Ti_2NiCu$ phase. This substitution seems to be corroborated by the sum of numbers of homopolar Ni-Ni and Cu-Cu pairs and by the sum of numbers of heteropolar Ti-Ni and Ti-Cu pairs, which are slightly greater than those of Ni-Ni and Ti-Ni pairs found in the amorphous $Ni_{46}Ti_{54}$ alloy [30].

It is known that the orientational correlations in disordered structures could be well represented by the distribution of the cosines of the bond-angles $\beta[\cos(\theta)]$. Bonds were defined by neighbors within the first coordination shell, considering the upper limit values used to calculate the coordination numbers. Figs. 13 and 14 show the $\Theta_{i\text{-}j\text{-}l}$ bond-angle distributions (the angle is centered at the middle atom $j$) obtained from the final atomic configuration for the input data set formed by the six $S(K)$ and Ni-$DSF(K)$ and Cu-$DSF(K)$ factors. The $\Theta_{i\text{-}j\text{-}l}$ bond-angle distributions obtained from the final atomic configuration and obtained for the second input data set (formed by four smoothed $S(K)$ factors) were similar. All the bond-angle distribution curves show peaks centered at about $\cos(\theta) \approx 0.601$ ($\theta = 53°$) and $\cos(\theta) \approx -0.243$ ($\theta = 104°$). The triangle and ideal tetrahedral angles are $\theta = 60°$ and $109.5°$, respectively. The $\Theta_{i\text{-}j\text{-}l}$ bond-angle distributions reported for the amorphous $Ni_{46}Ti_{54}$ alloy show peaks centered at about $\cos(\theta) \approx 0.50$ ($\theta = 60°$) and $\cos(\theta) \approx -0.375$ ($\theta = 112°$) [30]. By comparing the $\Theta_{i\text{-}j\text{-}l}$ bond-angle distributions for these two amorphous alloys it seems that the replacement of Ni by Cu in the amorphous $Ti_{50}Ni_{25}Cu_{25}$ structure promotes distortion in the chemical bonds, mainly in the Cu-Cu and Ti-Cu ones as suggested by the asymmetry seen in the first Cu-Cu and Ti-Cu neighbor shells displayed in Fig. 12.

## 6. Conclusions

A stoichiometric $Ti_{50}Ni_{25}Cu_{25}$ mixture was submitted to mechanical milling, and after 19 h of milling the XRD pattern showed the presence of an amorphous $Ti_{50}Ni_{25}Cu_{25}$ phase. Due to the experimental conditions present at the DB12A (XRD1) beamline during the AWAXS measurements, an approach to subtract the air scattering and Ni and Cu $K_\beta$ fluorescence signals from measured scattered intensities was developed. Two input data sets, the first consisting of six smoothed $S(K)$ and two smoothed $DFS(K)$ factors and the second consisting of four smoothed $S(K)$ factors, were used for the RMC simulations. Although the contributions of the $S_{Cu\text{-}Cu}(K)$ and $S_{Ni\text{-}Ni}(K)$ factors to $S(K)$ are very small, the $S_{ij}(K)$ factors and $G_{ij}(r)$ functions obtained from the RMC simulations considering both input data sets showed good agreement. The coordination numbers and interatomic distances for the first neighbors obtained for both input data sets also showed good agreement. The resuts suggest that in the amorphous $Ti_{50}Ni_{25}Cu_{25}$ phase the Cu atoms replace Ni ones as in the crystalline cubic $Ti_2NiCu$ phase, promoting distortions in the chemical bonds.

**Acknowledgments**

The authors thank the Brazilian agencies CNPq, CAPES, FAPESC and LNLS for financial support. We are indebted to the staff of the DB12A beamline (LNLS) for assistance during the AWAXS measurements.

TABLE

Table 1: $f'$ and $f''$ values used here.

| Energy (eV) | $f'_{Ti}$ | $f''_{Ti}$ | $f'_{Ni}$ | $f''_{Ni}$ | $f'_{Cu}$ | $f''_{Cu}$ |
|---|---|---|---|---|---|---|
| **8233** | 0.212 | 1.746 | -4.235 | 0.491 | -2.225 | 0.565 |
| **8333** | 0.229 | 1.697 | -7.714 | 1.407 | -2.371 | 0.551 |
| **8433** | 0.239 | 1.668 | -3.913 | 3.816 | -2.501 | 0.541 |
| **8877** | 0.278 | 1.534 | -2.051 | 3.503 | -3.934 | 0.495 |
| **8979** | 0.288 | 1.496 | -1.784 | 3.427 | -8.108 | 1.483 |
| **9077** | 0.292 | 1.477 | -1.668 | 3.387 | -4.384 | 3.843 |

FIGURES

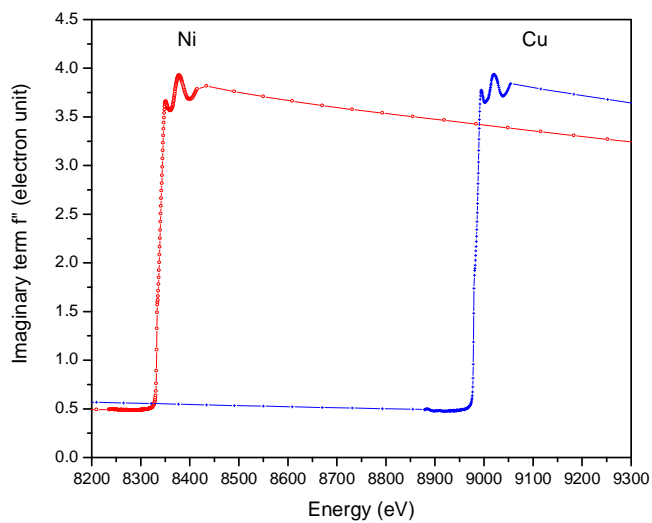

Fig. 1 (color online): Imaginary part $f''$ of the atomic scattering factor about the K edges of Ni and Cu in the amorphous $Ti_{50}Ni_{25}Cu_{25}$ alloy.

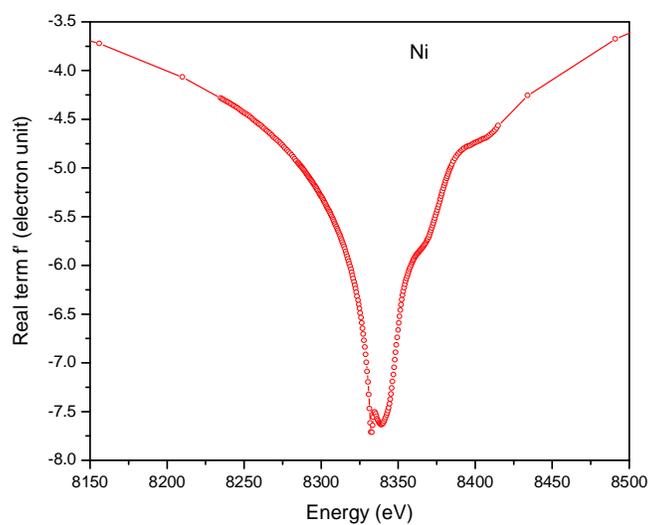

Fig. 2 (color online): Real part $f'$ of the atomic scattering factor about the K edge of Ni in the amorphous $Ti_{50}Ni_{25}Cu_{25}$ alloy.

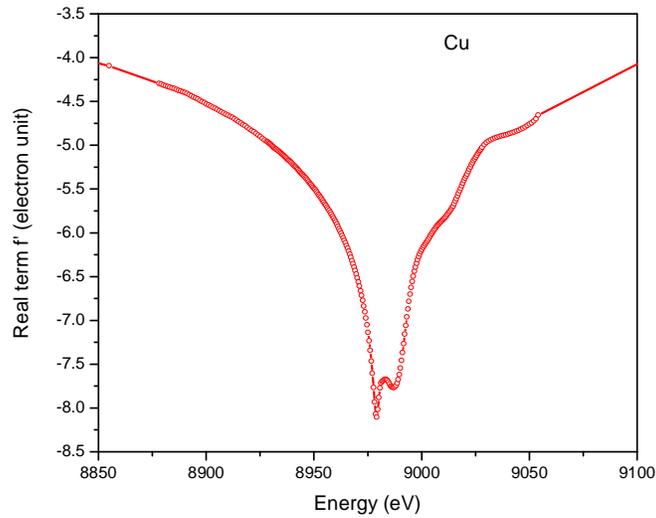

Fig. 3 (color online): Real part $f'$ of the atomic scattering factor about the K edge of Cu in the amorphous $Ti_{50}Ni_{25}Cu_{25}$ alloy.

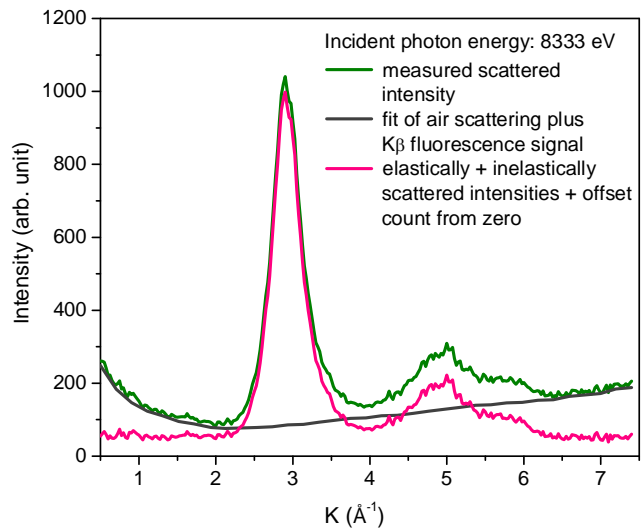

Fig. 4 (color online): Measured scattered intensity at the energy 8333 eV (dark olive solid curve), fit of air scattering +Ni$K_\beta$ fluorescence signal (dark gray solid curve), and elastically + inelastically scattered intensities + offset count from zero (pink solid curve).

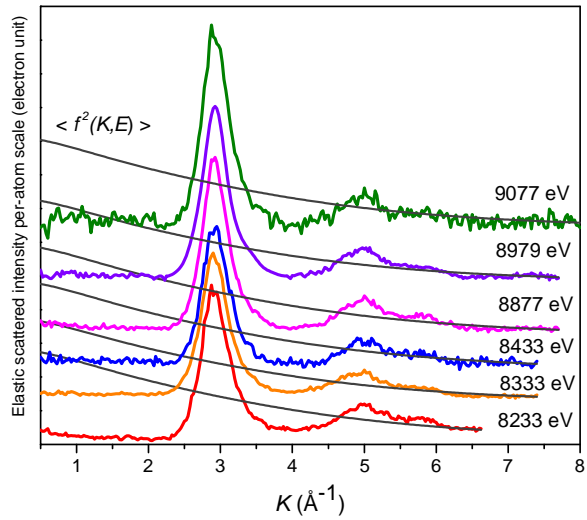

Fig. 5 (color online): Elastically scattered intensities on a per-atom scale together with the mean-square scattering factors (dark gray solid curves).

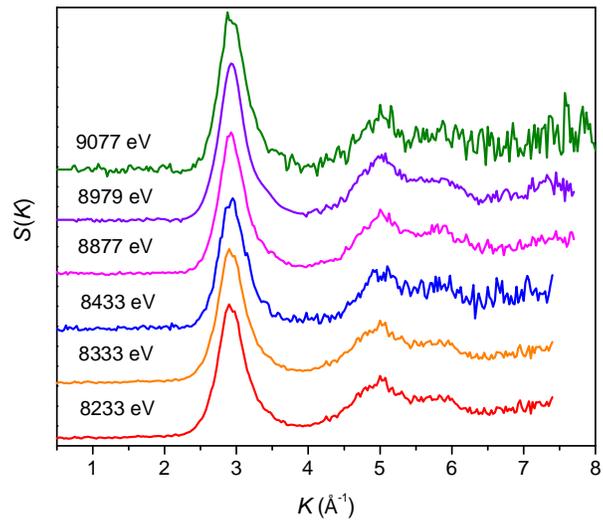

Fig. 6 (color online): Total structure factors $S(K)$ for the incident photon energies listed in Table 1.

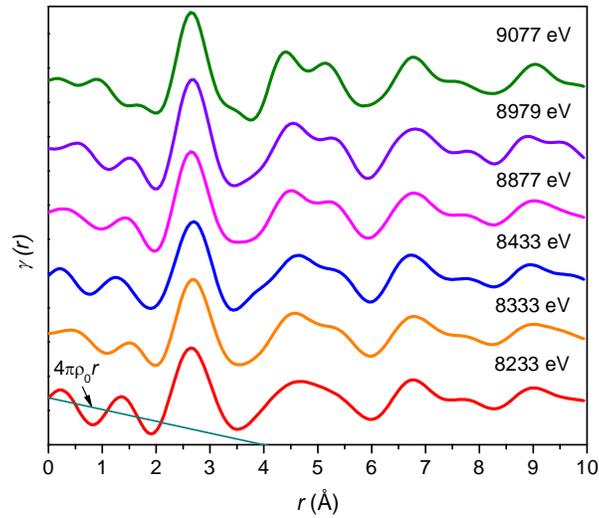

Fig. 7 (color online): Reduced total distribution function $\gamma(r)$ obtained from the Fourier transformation of $S(K)$ factors. The straight line $\gamma(r) = 4\pi\rho_0 r$ is also shown.

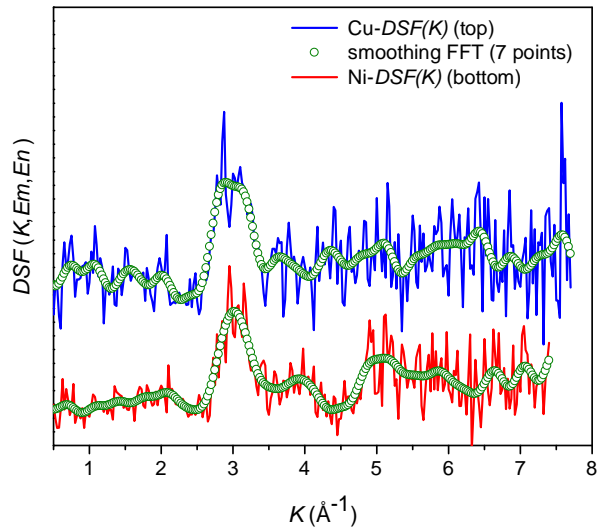

Fig. 8 (color online): Differential structure factors Ni-$DSF(K)$ and Cu-$DSF(K)$. The smoothed open circle curves were obtained after smoothing them using FFT filter tool, with 7 points, available in the Origin software.

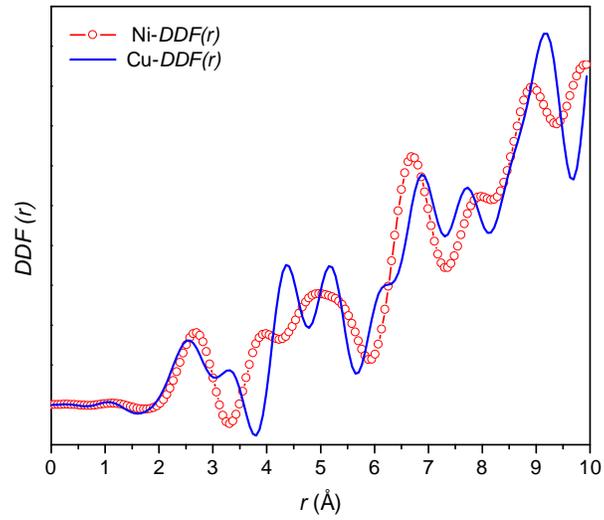

Fig. 9 (color online): Differential distribution functions *DDF*(*r*) obtained from the Fourier transformation of *DSF*(*K*) factors.

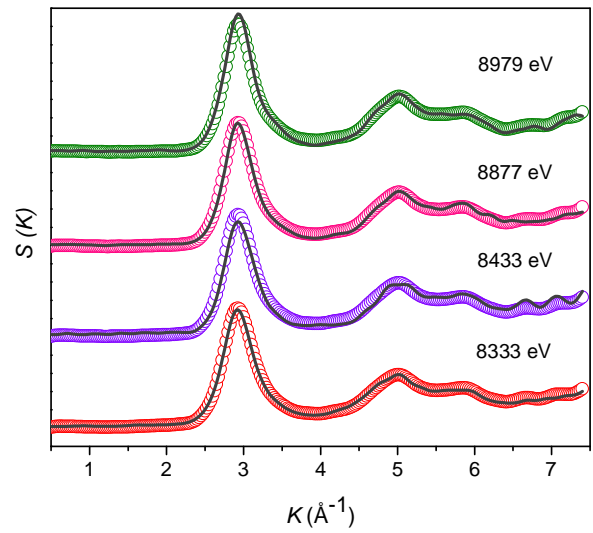

Fig. 10 (color online): Experimental (open circle curves) and simulated (dark gray solid curves) *S*(*K*) factors.

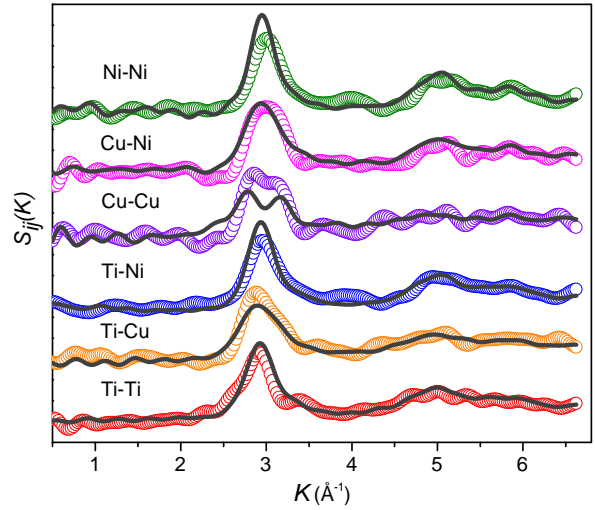

Fig. 11 (color online): Partial structure $S_{ij}(K)$ factors obtained from the RMC simulations using the first input data set (open circle curves) and the second one (dark gray solid curves). See the text for input data sets.

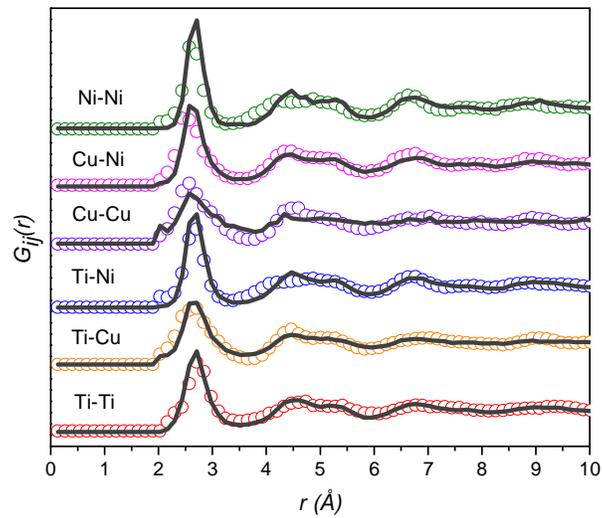

Fig. 12 (color online): Partial pair distribution functions $G_{ij}(r)$ obtained from the RMC simulations using the first input data set (open circle curves) and the second one (dark gray solid curves).

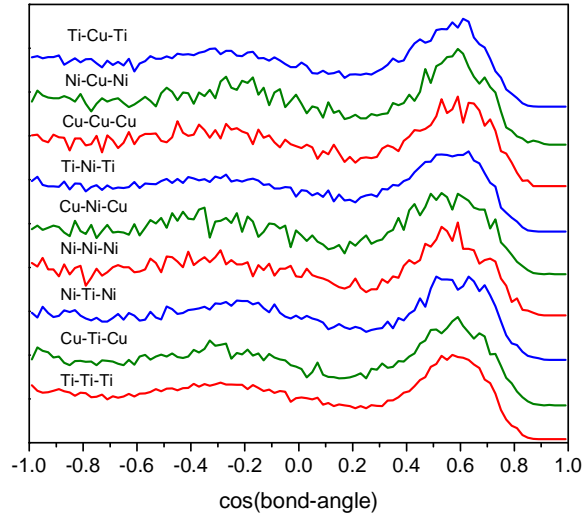

Fig. 13 (color online): Ti-Ti-Ti, Cu-Ti-Cu, Ni-Ti-Ni, Ni-Ni-Ni, Cu-Ni-Cu, Ti-Ni-Ti, Cu-Cu-Cu, Ni-Cu-Ni, Ti-Cu-Ti bond-angle distributions (the angle is centered at the middle atom) obtained from the final atomic configuration.

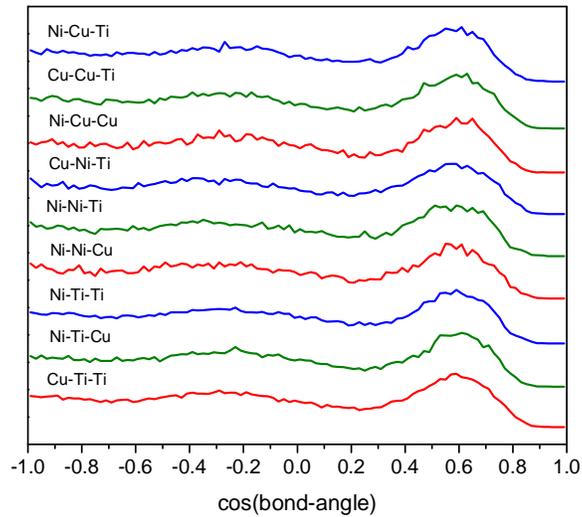

Fig. 14 (color online): Cu-Ti-Ti, Ni-Ti-Cu, Ni-Ti-Ti, Ni-Ni-Cu, Ni-Ni-Ti, Cu-Ni-Ti, Ni-Cu-Cu, Cu-Cu-Ti, Ni-Cu-Ti bond-angle distributions (the angle is centered at the middle atom) obtained from the final atomic configuration.